\documentstyle[12pt]{article}
\begin{document}

\title{High-spin states in boson models\\
with applications to actinide nuclei}

\author{S. Kuyucak$^1$ and S.C. Li \\
Department of Theoretical Physics,\\ Research School of Physical Sciences,\\
Australian National University, Canberra, ACT 0200, Australia}

\date{}

\maketitle

\begin{abstract}
We use the 1/$N$ expansion formalism in a systematic study of high-spin states
in the  $sd$ and $sdg$ boson models with emphasis on spin dependence of
moment of inertia and E2 transitions.
The results are applied to the high-spin states in the actinide nuclei
$^{232}$Th, $^{234-238}$U, where the need for $g$ bosons is especially acute
but until now, no realistic calculation existed. We find that the $d$-boson
energy plays a crucial role in description of the high-spin data.

\end{abstract}

\noindent
$^1$ E-mail: sek105@phys.anu.edu.au
\vfill \eject

\baselineskip=20pt

One of the early predictions of the the interacting boson model \cite{iac87}
(IBM) is the so called boson cutoff effect. Due to finite number of bosons
$N$ in the system, the ground band terminates at spin $L=l_{max}N$, where
$l_{max}$ denotes the maximum boson spin. As a result, yrast $B(E2)$ values
are predicted to fall off relative to the rotor values, vanishing at the
maximum
spin. In the $sd$-IBM, the cutoff occurs at $L=2N$ which is low enough to be
experimentally accessible and, therefore, has prompted many investigations
(see, for example, \cite{gro81,owe82,czo86,mau90,fah92}).
When the predicted falloffs did not materialize in any of the experiments
this was interpreted as due to insufficient collectivity of the $sd$-boson
system at high-spins which could be ameliorated by including $g$ bosons
\cite{raj82,wu82,aki87}.
Since then, evidence for $g$ bosons from low-spin spectra
have continuosly grown with concurrent developments in the $sdg$-IBM
calculations (see \cite{dev92,hey93} for recent reviews).
Ironically, the high-spin data, which have been the harbinger of $g$ bosons,
could not be treated in realistic $sdg$-IBM calculations. This happens because
the basis space for deformed nuclei ($N=12-16$) is too large and has to be
severely truncated for diagonalization, limiting their validity to low-lying
states.
Although the SU(3) limit has been used in discussing some features of high-spin
states in the $sdg$-IBM \cite{raj82,wu82,aki87}, realistic calculations
indicate that this limit is rather strongly broken \cite{cas88}.
Hence it can only provide a qualitative picture but otherwise is not very
useful
in applications to spectroscopy, or in addressing some basic questions on the
shortcomings of the $sd$-IBM raised by Bohr and Mottelson \cite{bm82},
e.g. the spin dependence of moment of inertia. An unfortunate consequence of
the lack of $sdg$-IBM calculations is that the $sd$-IBM results are still being
used in comparisons with high-spin data with obvious negative connotations (see
Refs. \cite{fah92} and \cite{tro94} for recent examples in experimental and
theoretical studies).

The angular momentum projected mean field theory, which
leads to a 1/$N$ expansion for physical quantities \cite{kuy88}, presumably
offers the only viable alternative for a realistic description of high-spin
states in the $sdg$-IBM. An obstacle in realizing this goal,
namely evaluation of matrix elements up to order $(L/N)^6$ which is necessary
for an accurate representation of high-spin states, has been recently overcome
through the use of computer algebra \cite{kuy95}.
The purpose of this letter is to use the extended 1/$N$ expansion formulas in a
systematic study of high-spin states (before the backbend) in boson models.
We address, in particular, the moment of inertia and boson cutoff problems in
the $sd$-IBM, and discuss how far they are resolved with the inclusion of $g$
bosons. Applications are made to the actinide nuclei, $^{232}$Th,
$^{234-238}$U, for which the ground band has been followed up to spins $L>2N$,
and hence the need for $g$ bosons is most acute.

The 1/$N$ expansion method has previously been discussed in detail \cite{kuy88}
and the recent extensions to higher orders are given in Ref.
\cite{kuy95}. Therefore, we give only a short account of the formalism
here.
The starting point of the $1/N$ calculations is the boson condensate
\begin{equation}
|N,{\bf x}\rangle =(N!)^{-1/2}(b^\dagger)^N|0\rangle,\quad
b^\dagger=\sum_l x_l b_{l0}^\dagger,
\end{equation}
where $b^\dagger_{l0}$ denote the boson creation operators and $x_l$
the associated mean fields. The summation index runs over $l=0,2$ in the $sd$
model and $l=0,2,4$ in the $sdg$ model. The axial symmetry assumption in Eq.
(1) has been justified in Ref. \cite{kuy95} by comparing the 1/N expansion
results with the exact diagonalization ones.
For a given Hamiltonian $H$, one evaluates the projected energies
$E_L=\langle N,{\bf x}| H P^L_{00} | N,{\bf x}\rangle$
and determines $x_l$ by variation after projection (VAP)
\cite{kuy88}.
The resulting energy expression is a double expansion in 1/$N$ and
$\bar L=L(L+1)$, and has the generic form
\begin{equation}
E_L = N^2 \sum_{n,m} {e_{nm}\over (aN)^m}
\Bigl({\bar L \over a^2N^2}\Bigr)^n,
\label{me1}
\end{equation}
where $a=\sum_l \bar l x_l^2$ and the expansion coefficients $e_{nm}$ involve
various quadratic forms of the mean fields $x_{l}$.
The coefficients $e_{nm}$ have recently been derived
up to the order $\bar L^3/N^6$ \cite{kuy95}. Since the $N$
dependence is not relevant to the discussion, we will suppress it for
simplicity and rewrite the energy formula (\ref{me1}) as
\begin{equation}
E_L=e_1 \bar L + e_2 \bar L^2 + e_3 \bar L^3,
\end{equation}
where the coefficients $e_n$ can be read off from the expressions given in
Ref. \cite{kuy95}. We note that similar expressions are used in the geometrical
model analysis of deformed nuclei \cite{bm75}. The difference between the two
models is that in the IBM the coefficients $e_n$ follow from an underlying
Hamiltonian (which is used in describing other properties) whereas in the
geometrical model they are directly extracted from the data. The moment of
inertia problem raised in Ref. \cite{bm82} refers to the fact that in the
$sd$-IBM, i) the $e_1$ coefficient gets a substantial contribution from the
dipole interaction, $L\cdot L$, which has no dynamical content, and ii) the
$e_2$ coefficient is much smaller than the experimental values.
The two problems are in fact interrelated. Although the latter can be resolved
by renormalizing the moment of inertia at high-spins (e.g. by modifying
$L\cdot L \to L\cdot L/(1+f L\cdot L$) \cite{yos91}), such modifications are
purely kinematical in origin and do not address the dynamical problem.

Another observable of interest in the study of high-spin states is the yrast
$E2$ transitions which have the generic form
\begin{equation}
\langle L+2 \parallel T(E2) \parallel L \rangle =
\alpha N \hat L \langle L0\, 20|L+2\ 0\rangle \bigl[m_1 + m_2 L(L+3)\bigr]
\label{e2}
\end{equation}
where $\alpha$ is an effective boson charge, $\hat L = [2L+1]^{1/2}$
and the coefficients $m_n$ are given in Ref. \cite{kuy95}. The first term in
(\ref{e2}) gives the familiar rigid-rotor result. The second term is negative
and is responsible for the falloffs predicted in $E2$ transitions.

In order to compare the $sd$ and $sdg$ model predictions,
we first present a brief study of the $sd$-IBM results. We use the standard
Hamiltonian
\begin{equation}
H=-\kappa Q \cdot Q  + \kappa' L \cdot L + \varepsilon_d n_d ,
\label{ham}
\end{equation}
where $L$ and $n_d$ are the angular momentum and $d$-boson number operators,
and the quadrupole operator is given by
\begin{equation}
Q=[s^\dagger {\tilde d} + d^\dagger {\tilde s}]^{(2)} +
\chi [d^\dagger {\tilde d}]^{(2)}.
\label{qsd}
\end{equation}
Here brackets denote tensor coupling of the boson operators and
$\tilde b_{lm}=(-1)^{m}b_{l-m}$.
For consistency, the same quadrupole operator is used in the $E2$ transition
operator, $T(E2)=\alpha Q$ (\ref{e2}) as in the Hamiltonian (\ref{ham}).
Since $L \cdot L$ does not play any role in the dynamics of the system, we will
not discuss it further (it can be easily restored by changing $e_1 \to
e_1+\kappa'$). In presenting systematics, we find it convenient to use ratios
which eliminate the undesired effects of the scale parameters $\kappa$ and $N$.
The energy scale can be fixed, for example, by fitting $\kappa$ to the
excitation energy of the $\gamma$ band, $E_\gamma$.
In Fig. 1, we show four such quantities as a function of $q=\chi/\chi_{SU3}$
for various values of $\eta_d=\varepsilon_d/N \kappa$. We comment on their
behaviour and contrast them with the experimental data below. \\
a) $E_\gamma/N e_1$: This ratio relates the energy scales of the $\gamma$ and
ground bands, and its mismatch with experiment has been a source of criticism
\cite{bm82}. It is around 4-5 in the rare-earth region and
increases to 8-10 in the actinides. The SU(3) limit ($q=1, \eta_d=0$) is
seen to give the maximum value which overestimates it by a factor of 2-4.
It decreases rapidly with $\varepsilon_d$ and $q$ though, and through a
judicious use of these parameters, it should be possible to describe this ratio
(and hence the the moment of inertia) without using the $L\cdot L$ term. \\
b) $N^2 e_2/e_1$: This ratio measures the deviation from the rigid rotor
behaviour (SU(3) limit) due to loss of pairing. It ranges from about -0.2 in
the rare-earth region to -0.1 in the actinides. Clearly, it can not be
explained by the standard $sd$-IBM Hamiltonians currently in use for deformed
nuclei which assume $\eta_d=0$, $q\sim 0.4-0.5$.
However, it is quite sensitive to $\eta_d$ values and the experimental range
can be easily attained by including the $d$-boson energy in the Hamiltonian. \\
c) $N^4 e_3/e_1$: There is some uncertainty in the extraction of this ratio
from data, especially in the rare-earth region. In the actinides, it is about
0.01. It depicts even more sensitivity to both $q$ and $\eta_d$, and therefore
its description should not pose any problems. \\
d) $N^2 m_2/m_1$: As there is no boson cutoff effect, experimentally this ratio
is consistent with zero. For $\eta_d=0$, it remains rather flat at the
SU(3) value which gives the maximum possible effect. Introduction of one-body
energy, however, reduces it substantially, becoming more in line with
experiments.

The $d$-boson energy has been mostly neglected in studies of deformed nuclei,
presumably due to the success of the consistent-Q formalism (variable $\chi$
with $\varepsilon_d=0$) in explaining the energy and $E2$ transition
systematics of low-lying states \cite{cas88}. In fact, for small values
($\eta_d\sim 1)$, its effect on low-lying states is negligible and it is
not really needed in their description \cite{lip85}.
The above analysis indicates that breaking of the SU(3) limit by either the
pairing interaction \cite{bm82} or by varying the $\chi$ parameter \cite{cas88}
does not lead to a soft enough energy surface which is the main reason for the
perceived moment of inertia problem in the $sd$-IBM. The obvious way towards a
softer energy surface is to include the $d$-boson energy in the Hamiltonian
which is seen to vastly improve the description of the spin-dependent terms in
the ground energies and $E2$ transitions.

We next present a similar study in the $sdg$-IBM.
A minimal extension of the $sd$-IBM to the $sdg$ model can be achieved by
including the $g$-boson energy term, $\varepsilon_g n_g$ in the
Hamiltonian (\ref{ham}), and modifying the quadrupole operator (\ref{qsd}) to
\begin{equation}
Q=[s^\dagger {\tilde d} + d^\dagger {\tilde s}]^{(2)} +
q_{22} [d^\dagger {\tilde d}]^{(2)} + q_{24} [d^\dagger {\tilde g} +
g^\dagger {\tilde d}]^{(2)} + q_{44} [g^\dagger {\tilde g}]^{(2)}.
\label{qsdg}
\end{equation}
The quadrupole parameters $q_{24}$ and $q_{44}$ play an important role in the
description of the hexadecapole bands but otherwise, the ground band properties
are not very sensitive to their variations.
In order to limit the number of parameters, we scale the three quadrupole
parameters $q_{22},q_{24},q_{44}$ from their SU(3) values with a single factor
$q$ as suggested by microscopics \cite{oai78}.
In Fig. 2, we show the ratios in Fig. 1 as a function of $q$
for various values of $\eta_g=\varepsilon_g/N \kappa$ with $\eta_d=0$.
Before commenting on specific ratios, we point out some general features.
For large $\eta_g$, the $g$ bosons decouple and the results converge to those
of the $sd$ model shown in Fig. 1. This convergence is apparent from the
overlap of lines in Figs. 2-a and b but requires even larger values of $\eta_g$
in the case of 2-c and d. It is harder to pin down realistic values for
$\eta_g$ due to lack of data, nevertheless, we quote the literature values
for comparison which range from 3-6. \\
a) $E_\gamma/N e_1$: Inclusion of $g$ bosons increases this ratio which is
contrary  to the experimental trend. However, for realistic $\eta_g$ values,
this adverse change is too small to worry about. \\
b) $N^2 e_2/e_1$: This ratio also increases (in absolute value) which is good
but again too small for realistic $\eta_g$ values to have any impact. \\
c) $N^4 e_3/e_1$: This ratio shows some sensitivity to $g$ bosons, however,
it is nowhere near the effect of $\eta_d$ in Fig. 1-c, and therefore
not likely to have much relevance. \\
d) $N^2 m_2/m_1$: The boson cutoff was the original reason for the introduction
of $g$ bosons and it is clear from this figure why. In the SU(3) limit,
this ratio is reduced by a factor of 4 compared to the $sd$-IBM. Its $q$ and
$\eta_g$ dependence, however, is opposite to that of Fig. 1-d, and things
get worse away from the SU(3) limit. For realistic parameters, the reduction
from the $sd$-IBM result (with $\eta_d=0$) is less than 40\% which is certainly
not enough, and one needs the $d$-boson energy to reduce it further.

The somewhat surprising conclusion of the above systematic study is that
introduction of the $g$ bosons, though necessary to describe states with
$L>2N$,
hardly improves the dynamics of the boson system. The problems attributed
to the $sd$-IBM are, in fact, due to not having a soft enough energy surface
and can only be resolved by including the $d$-boson energy in the Hamiltonian
(and not by introduction of $g$ bosons alone).

In the light of the systematic trends discussed above, we carry out fits
to the actinide nuclei, $^{232}$Th, $^{234-238}$U, for which the ground band
has been followed up to spins $L>2N$, and hence the need for $g$ bosons is most
acute but no realistic IBM calculation yet exists.
The parameters used in the fits are shown in Table 1. The ground band energies
(normalized with $L(L+1)$ for a fairer representation of all spins) are shown
in Fig. 3. A good description of the ground band energies is obtained
for all four nuclei with relative errors of about 1\%. It is worth emphasizing
that the moment of inertia and its spin dependence are explained without
appealing to the $L\cdot L$ term or its phenomenological variations.
To appreciate the importance of the $d$-boson energy in the fits, we note from
Fig. 2b that with $\eta_d=0$, the slopes of the curves in Fig. 3 would be an
order of magnitude smaller, leading to almost flat lines like in a rigid rotor.
In Fig. 4, the calculated $E2$ transitions along the ground band are compared
to experiment. Again a reasonably good description of the data is obtained.
Finally, to show that the ground band results are not achieved at the expense
of other bands,  we compare in Table 2 the predictions for the $\beta, \gamma$
band head energies and  $E2(2_{\beta, \gamma} \to 0_g)$ matrix elements with
experiment. Note that these calculations are the leading order $1/N$ expansion
results and hence are correct to order $1/N$. The good agreement obtained in
Table 2 verifies that the inclusion of the $d$-boson energy does not detract
from the usual quality of the $\beta$ and $\gamma$ band systematics achieved
in the IBM calculations.

In conclusion, we reiterate that the perceived problems with the $sd$-IBM in
its description of spin dependent quantities is not due to lack of higher spin
bosons but rather due to the energy surface not being soft enough. Inclusion of
the $d$-boson energy, together with the extension to $sdg$ space, can
successfully resolve these problems as demonstrated in the fits to the actinide
nuclei. The $g$ bosons are necessary for extending the model space but
otherwise they play a marginal role in the dynamics of the ground band
and can not resolve alone the problems mentioned above.
Details of the present work, including applications to the rare-earth nuclei,
will be presented in a longer paper \cite{kuy96}.

This research was supported by the Australian Research Council.
S.K. thanks Prof. von Brentano for useful discussions and the members of the
IKP at the University of K\"oln for their hospitality.

\vfill \eject

\vfill \eject

\begin{table}
\caption{Parameters used in the $sdg$-IBM calculations. $\kappa$ is in keV and
$\alpha$ in eb.}
\label{table1}
\begin{tabular}{ccccccc} \hline
Nucleus & $N$ &  $\kappa$ & $q$ & $\eta_d$ & $\eta_g$ & $\alpha$
\\ \hline
$^{232}$Th& 12 & 14.3 & 0.7 & 1.85 & 4.2 & 0.210 \\
$^{234}$U & 13 & 15.6 & 0.7 & 1.62 & 3.1 & 0.195 \\
$^{236}$U & 14 & 16.2 & 0.7 & 1.62 & 3.1 & 0.195 \\
$^{238}$U & 15 & 16.3 & 0.7 & 1.67 & 3.1 & 0.195 \\  \hline
\end{tabular}
\end {table}

\begin{table}
\caption{Comparison of the $\beta$ and $\gamma$ band energies (in keV) and $E2$
transitions (in eb) with the $sdg$-IBM calculations in the actinide region.
The data are from \protect \cite{nds}.}
\label{table2}
\begin{tabular}{cccccccccccccc} \hline
 &&\multicolumn{2}{c}{$E_\beta$}
&&\multicolumn{2}{c}{$E_\gamma$}
&&\multicolumn{2}{c}{$<2_\beta ||T(E2)||0_g>$}
&&\multicolumn{2}{c}{$<2_\gamma ||T(E2)||0_g>$} \\[0.1cm]
\cline{3-4} \cline{6-7} \cline{9-10} \cline{12-13}
Nucleus &&\multicolumn{1}{c}{Cal.}&\multicolumn{1}{c}{Exp.}&&
   \multicolumn{1}{c}{Cal.}&\multicolumn{1}{c}{Exp.}&&
   \multicolumn{1}{c}{Cal.}&\multicolumn{1}{c}{Exp.}&&
   \multicolumn{1}{c}{Cal.}&\multicolumn{1}{c}{Exp.}\\
\hline
$^{232}$Th&& 731 & 730 && 786 & 785 && 0.27 & 0.33 $\pm$ 0.07 && 0.40 & 0.37
$\pm$ 0.07 \\
$^{234}$U && 804 & 809 && 872 & 927 && 0.21 & $<$0.25 && 0.35 & 0.36$\pm$0.08\\
$^{236}$U && 903 & 919 && 978 & 958 && 0.22 & - && 0.37 & - \\
$^{238}$U && 949 & 993 && 1040 & 1060 && 0.22 & 0.24$\pm$ 0.05&& 0.37 & 0.37
$\pm$ 0.04 \\  \hline
\end{tabular}
\end {table}

\vglue 10cm
\vfill \eject

{\Large \bf Figure captions}
\\[.5cm]
Fig. 1. Sytematic study of moment of inertia (a), its spin dependence (c,d),
and the boson cutoff effect in $E2$ transitions (d) in the $sd$-IBM.
The quadrupole parameter $q$ is normalized to 1 in the SU(3) limit, and
the $d$-boson energy parameter $\eta_d = \varepsilon_d/N \kappa$ is varied from
0-2 in 10 equal steps.
\\[.4cm]
Fig. 2. Same as Fig. 1 but in the $sdg$-IBM. The quadrupole parameters
$q_{22},q_{24}, q_{44}$ are scaled from their SU(3) values with a single factor
$q$, $\eta_d=0$ and the $g$-boson energy parameter $\eta_g = \varepsilon_g/N
\kappa$ is varied from 0-10 in 20 equal steps.
\\[.4cm]
Fig. 3. Comparison of the experimental (circles) and calculated (solid lines)
ground band energies $E_L/L(L+1)$ (in keV) in the actinide nuclei.
The data are from \cite{nds}.
\\[.4cm]
Fig. 4. Comparison of the experimental (circles) and calculated (solid lines)
yrast $E2$ transitions in the actinide nuclei. The data are from
\cite{owe82,nds}.

\end{document}